# The Deformation of Wrinkled Graphene

*Zheling Li[1], Ian A. Kinloch[1], Robert J. Young[1]\*, Kostya S. Novoselov[2], George Anagnostopoulos[3], John Parthenios[3], Costas Galiotis[3,4], Konstantinos Papagelis[3,5], Ching-Yu Lu[6], Liam Britnell[6]*

[1]School of Materials and [2]School of Physics and Astronomy, University of Manchester, Oxford Road, Manchester, M13 9PL, UK

[3]Institute of Chemical Engineering Sciences, Foundation for Research and Technology – Hellas (FORTH/ ICE-HT), P.O. Box 1414, Patras 26504, Greece,

[4]Department of Chemical Engineering and [5]Department of Materials Science, University of Patras, Patras 26504, Greece

[6]BGT Materials Ltd, Photon Science Institute, University of Manchester, Oxford Road, Manchester, M13 9PL, UK

**ABSTRACT**

The deformation of monolayer graphene, produced by chemical vapor deposition (CVD), on a polyester film substrate has been investigated through the use of Raman spectroscopy. It has been found that the microstructure of the CVD graphene consists of a hexagonal array of islands of flat monolayer graphene separated by wrinkled material. During deformation, it was found that the rate of shift of the Raman 2D band wavenumber per unit strain was less than 25% of that of flat flakes of mechanically-exfoliated graphene, whereas the rate of band broadening per unit strain was about 75% of that of the exfoliated material. This unusual deformation behavior has been modeled in terms of mechanically-isolated graphene islands separated by the graphene wrinkles, with the strain distribution in each graphene island determined using shear lag analysis. The effect of the size and position of the Raman laser beam spot has also been incorporated in the model. The predictions fit well with the behavior observed experimentally for the Raman band shifts and broadening of the wrinkled CVD graphene. The effect of wrinkles upon the efficiency of graphene to reinforce nanocomposites is also discussed.

Following its first isolation in 2004, graphene has shown huge potential in both fundamental studies[1] and industrial applications.[2] Currently one of the urgent targets is to grow large size, continuous and defect-free graphene. The chemical vapor deposition method (CVD) opens a route to achieve these targets at low cost.[3] However, the grain boundaries formed after the graphene grains become stitched together[4] affect its performance,[5-7] for example, in applications such as transparent electrodes since these boundaries impede electrical transport.[4,8] Moreover, CVD grown graphene typically has to be transferred to other substrates for use,[9,10] during which wrinkles can be induced[10,11] as a result of the different thermal expansion of the substrates,[12,13] the replication of the substrate topography,[14] and the transfer process itself.[9] These wrinkles have been observed on graphene-based transparent electrodes,[15,16] and are thought to further alter its mechanical stretchability,[17] electronic structure,[18] and also local potential.[19] It is also thought that the presence of wrinkles can affect the deformation of graphene in shear,[20] the deformation of graphene oxide paper[21] and the ability of graphene oxide to reinforce polymer matrices in nanocomposites.[22] Although wrinkling appears to be an inherent property of graphene due to its extremely low bending rigidity[23] even when it is fully embedded into polymer matrices,[24] there has as yet been no systematic experimental study of its effect upon the mechanical response of graphene.

Raman spectroscopy is a versatile tool to study graphene,[25-27] in particular, it can be useful in the study of number of layers,[25, 28] stacking order[28, 29] and doping etc.[30, 31] It has also been used to monitor the deformation of graphene,[32] and to demonstrate that continuum mechanics is still valid even for a one-atom thick material.[24, 33-35] In many of the previous studies on large flat graphene flakes, it was possible to assume uniformity of the graphene within the spatial resolution of the Raman laser spot (around 1-2 μm in diameter). However, when the graphene is <1 μm in size or at edges, Raman spectroscopy only provides limited information as a result of non-uniformity of the Raman scattering due to the structural non-uniformity of the graphene, and also the spatial variation of the excitation intensity over the laser spot.[36, 37] In fact when difficulties are encountered in this situation, it is usually assumed that the signal detected is the average Raman scattering emanating from all of the graphene within the spot.[24, 37, 38]

In this work, Raman spectroscopy has been employed to monitor the deformation mechanics of monolayer CVD graphene on a poly(ethylene terephthalate) substrate (CVD graphene/PET) where the PET film is flat but the graphene is wrinkled. It is found that upon deformation of the film, the shift of the graphene Raman 2D band with strain and the band broadening characteristics are quite different from that observed for mechanically-exfoliated monolayer graphene flakes. It is shown that the wrinkles have the effect of separating the graphene mechanically into isolated islands, with each island being similar in size to the Raman laser spot. It is demonstrated that inside each island the stress will be transferred non-uniformly from the PET to the graphene and this allows the unusual Raman band shift and broadening behavior to be explained. The effect of such wrinkles upon the ability of graphene to reinforce nanocomposites will also be discussed below.

**RESULTS AND DISCUSSION**

The scanning electron microscope (SEM) images in Figure 1(a) of the surface of the CVD graphene/PET show the network of CVD graphene islands separated by wrinkles with a height of around 20 nm, as revealed by atomic force microscopy (AFM) (Figures 1(b) and 1(d)).

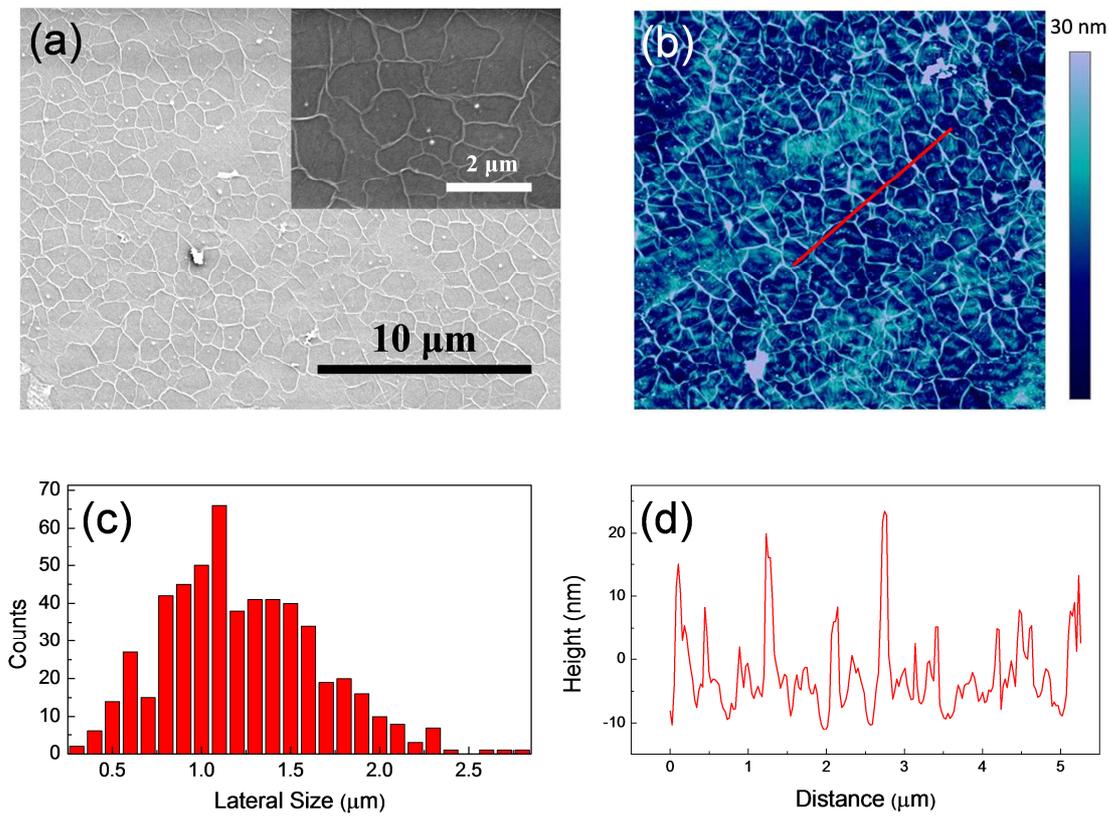

**Figure 1.** (a) SEM and (b) AFM images of the CVD graphene. (c) The distribution of the lateral dimensions of the graphene islands. (d) The AFM height profile of the inset red line in (b) showing the height of the wrinkles.

The wrinkled graphene microstructure resembles those found previously.[39, 40] It is thought that the wrinkles form in the CVD graphene for at least two reasons. First of all it appears that the Cu substrate employed is never completely flat. Secondly, they will form as a result of the different thermal coefficients of Cu substrate ($\approx 20 \times 10^{-6}$ K$^{-1}$)[41] and graphene ($-8.0 \times 10^{-6}$ K$^{-1}$).[42] After growth, when the CVD graphene cools from a typical growth temperature of 1000 °C, the Cu contracts but the graphene expands, resulting in a compressive strain of 2~3% in the graphene.[14] This is at least an order of magnitude higher than the critical strain required for graphene buckling.[24, 43] Such high strains can induce severe high order buckling deformation which does not relax back when the strain is released and gives rise to large a network of large out-of-plane wrinkles or folds, in good agreement with computational simulations.[18] It also bears a strong similarity to the wrinkled microstructure found for thin films of copper on layered-crystal surfaces, again formed through a mismatch in thermal expansion coefficients.[13]

The wrinkles separate the graphene surface into small isolated islands with their size distribution, based upon more than 500 measurements as detailed in the Supporting Information, shown in Figure 1(c). It can be seen that there is a broad distribution of the lateral dimensions of the islands, with a mean value around 1.2 μm, but also with some larger islands of up to 3 μm in diameter.

Raman spectroscopy has been employed to monitor interfacial stress transfer from the PET substrate to the CVD graphene and the whole CVD graphene/PET film has been modeled as a nanocomposite structure.[22, 33, 44] The Raman spectra of the PET substrate and the CVD graphene/PET are shown in Figure 2. The 2D band at around 2700 cm$^{-1}$ (also known as the G' band) results from two phonons with opposite momentum in the highest optical branch near the **K** point.[25] The graphene G band overlaps partially with the PET band (Figure 2) and so only the 2D band has been used here for the analysis of stress transfer. The lack of a visible D band suggests the absence of defects (grain boundaries, etc) even at the wrinkles.[4] An estimate of the intensity ratio of the 2D band to the G band (after deconvolution from the strong adjacent PET band), along with a sharp 2D band with a full width at half maximum around 30 cm$^{-1}$, demonstrates that the CVD graphene is essentially a monolayer.[25, 45]

The initial position of the 2D band of the CVD graphene on the PET film of 2696.0 ± 2.2 cm$^{-1}$ compared with the stress-free value of ~2677 cm$^{-1}$ from mechanically exfoliated monolayer graphene with 514 nm laser excitation[46] clearly indicates that graphene on the substrate is in compression. It appears that the graphene islands are able to support the compressive loads and it is only at the island boundaries that the compressive load is relieved by wrinkling.

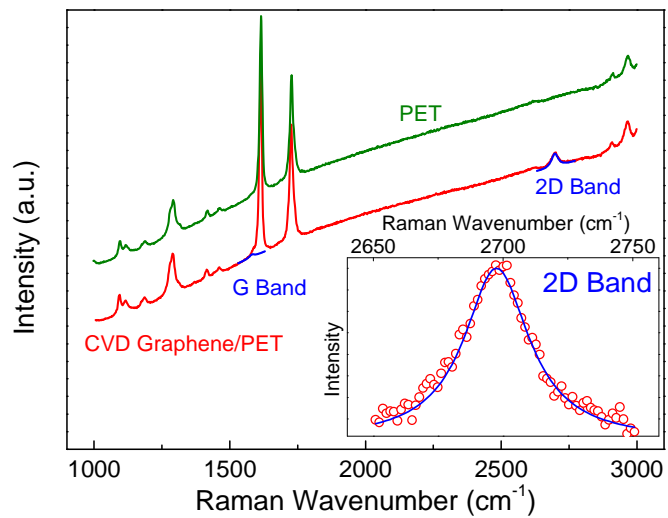

**Figure 2.** Raman spectra of the CVD graphene/PET and neat PET substrate. Inset shows the experimental data (red circle) and the Lorentzian fitting (blue line) for the 2D band. (The background scattering was not removed from the spectra)

The CVD graphene/PET was subjected to tensile deformation and the shift of the graphene Raman 2D band, fitted with a Lorentzian function, was monitored to elucidate the deformation mechanics.[47-49] When the specimen was strained uniaxially in tension the 2D band position ($\omega_{2D}$) downshifted with strain $\varepsilon$ at a rate $d\omega_{2D}/d\varepsilon$ = -12.8±2.0 cm$^{-1}$/%. At the same time the 2D band broadened with strain and its full width at half maximum ($FWHM_{2D}$) increased approximately linearly with $\varepsilon$ at a rate of $dFWHM_{2D}/d\varepsilon$ = 9.3±3.1 cm$^{-1}$/% (Figure 3). The absence of discontinuities in the data in Figure 3 implies that the interface between the CVD graphene and the PET film remains intact up to a strain of at least 0.4%.[50] The data shown in Figure 3 are determined from separate sets of measurements from 8 regions chosen a random the CVD graphene/PET as shown in the Supporting Information. The data points in Figure 3 were averaged from the 8 sets of measurements at each strain level and the standard deviations given above are determined from the slopes of the individual sets of measurements in the Supporting Information. The scatter in the data is a reflection of the variation in size of the individual graphene islands as shown in Figure 1(c).

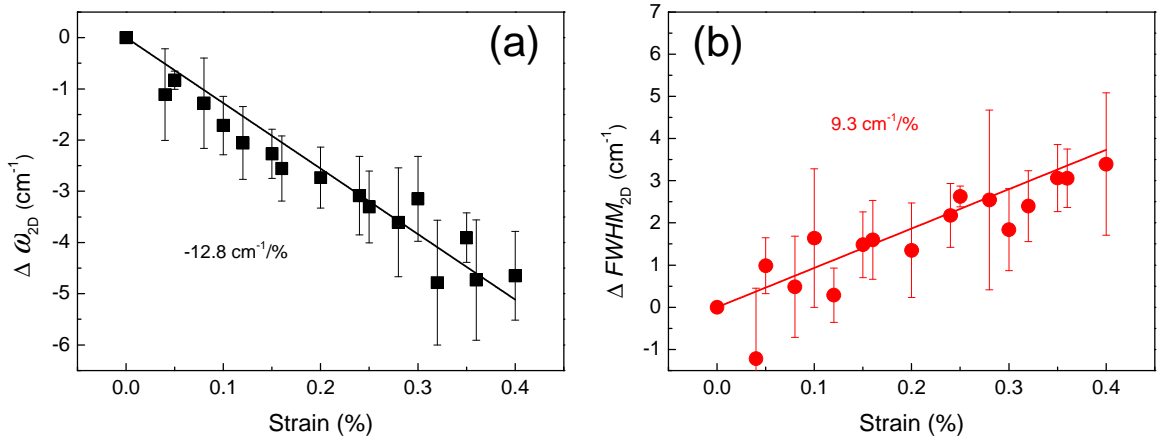

**Figure 3.** The variation of (a) $\omega_{2D}$ and (b) $FWHM_{2D}$ under uniaxial strain. The red lines are linear fits for both sets of data (mean values of 8 sets of measurements).

Generally tensile strain induces phonon softening in graphene,[48] which can be estimated using the knowledge of the Grüneisen parameter.[51-53] For an ideal flat monolayer of graphene under uniaxial strain,[52] the reference 2D band shift rate $(d\omega_{2D}/d\varepsilon)_{ref}$ is given as:

$$\left(\frac{d\omega_{2D}}{d\varepsilon}\right)_{ref} = -\omega_{2D}^0 \gamma_{2D}(1-v) \quad (1)$$

where $\omega_{2D}^0$ is the $\omega_{2D}$ at zero strain, $\gamma_{2D}$ is the Grüneisen parameter for 2D band, and $v$ is the Poisson's ratio of the substrate. As reported recently,[54] the value of $(d\omega_{2D}/d\varepsilon)_{ref}$ is dependent on the laser excitation and the Poisson's ratio of the matrix. In this study, if $\gamma_{2D} = 3.55$ and $v = 0.35$ is taken for PET,[52] the value of $(d\omega_{2D}/d\varepsilon)_{ref}$ is estimated to be around -60 cm$^{-1}$/% for flat monolayer on PET substrate. Additionally, the reference 2D band broadening rate with $\varepsilon$ (($dFWHM_{2D}/d\varepsilon)_{ref}$) is found experimentally to be ~12 cm$^{-1}$/% using the 514 nm laser excitation.[34] That is to say, when the monolayer graphene on PET substrate is fully stretched to a strain of 1%, $\omega_{2D}$ downshifts by 60 cm$^{-1}$ while at the same time $FWHM_{2D}$ increases by 12 cm$^{-1}$. Hence, in the present study (Figure 3), the measured value of 2D band shift $d\omega_{2D}/d\varepsilon$ is less than 25% of $(d\omega_{2D}/d\varepsilon)_{ref}$, while the broadening rate $dFWHM_{2D}/d\varepsilon$ is nearly 75% of that of $(dFWHM_{2D}/d\varepsilon)_{ref}$.[34, 52, 55] Such anomalous behavior is not found during the deformation of flat mechanically-exfoliated monolayer graphene on PET[56] but wrinkles are invariably found when CVD graphene is transferred to a polymer substrate.[10, 57] In a previous study[58] we measured a band shift rate $d\omega_{2D}/d\varepsilon$ of around -30 cm$^{-1}$/% strain for CVD graphene deformed on a poly(methyl methacrylate) substrate, in which there were fewer wrinkles present than in the material employed in this present study.

It will now be demonstrated that the low band shift rate $d\omega_{2D}/d\varepsilon$ and higher-than-expected rate of broadening $dFWHM_{2D}/d\varepsilon$ during deformation are both the result of the CVD graphene monolayer being wrinkled. In order to model this deformation behavior, it has been assumed that the microstructure of the graphene consists of a series of graphene islands bonded to the PET substrate, averaging 1.2 μm in diameter, separated by wrinkles that do not allow the transfer of stress between the isolated islands as shown in Figure 4(a). Hence, in each individual island the axial strain will build up from zero at the wrinkles to become a maximum in the middle of the island. The exact form of the strain distribution will depend upon the size of the island and the efficiency of stress transfer from the underlying PET substrate. Figure S2 in the Supporting Information shows schematically how it is envisaged that the presence of the wrinkles leads to the graphene islands being isolated mechanically.

It is then assumed that the ~1.2 μm diameter graphene islands can be modeled as 12 strips of ~0.1 μm wide mechanically-independent graphene nanoribbons lying parallel to the direction of tensile stress (Figure 4(b)). Furthermore, it is assumed that the strain distributions in each nanoribbon can be estimated using 'shear-lag theory' as has been done earlier for exfoliated graphene flakes subjected to deformation on a polymer substrate.[33, 59] Because of the comparable size of the graphene island to the Raman laser spot, it is also necessary to consider the effect of laser spot size in the analysis.[36] It is shown in the Supporting Information that the effective size of the laser spot is of the order of 1.4 μm, i.e. similar to that of the graphene islands. Moreover, there is a variation of intensity across the spot (Gaussian distribution) as shown in the Supporting Information.[60] Considering the size of the wrinkles, it is reasonable to assume that most of the laser spot intensity (~90 %) is within the graphene islands. Hence in order to calculate the spatial distribution of local strain and laser beam intensity, each strip is further divided into ~0.1 μm square elementary units with their coordinates given by the longitudinal ($L$) and transverse ($T$) position parameters, where (-6 ≤ $L$ ≤ 6) and (-6 ≤ $T$ ≤ 6) (Figure 4(c)).

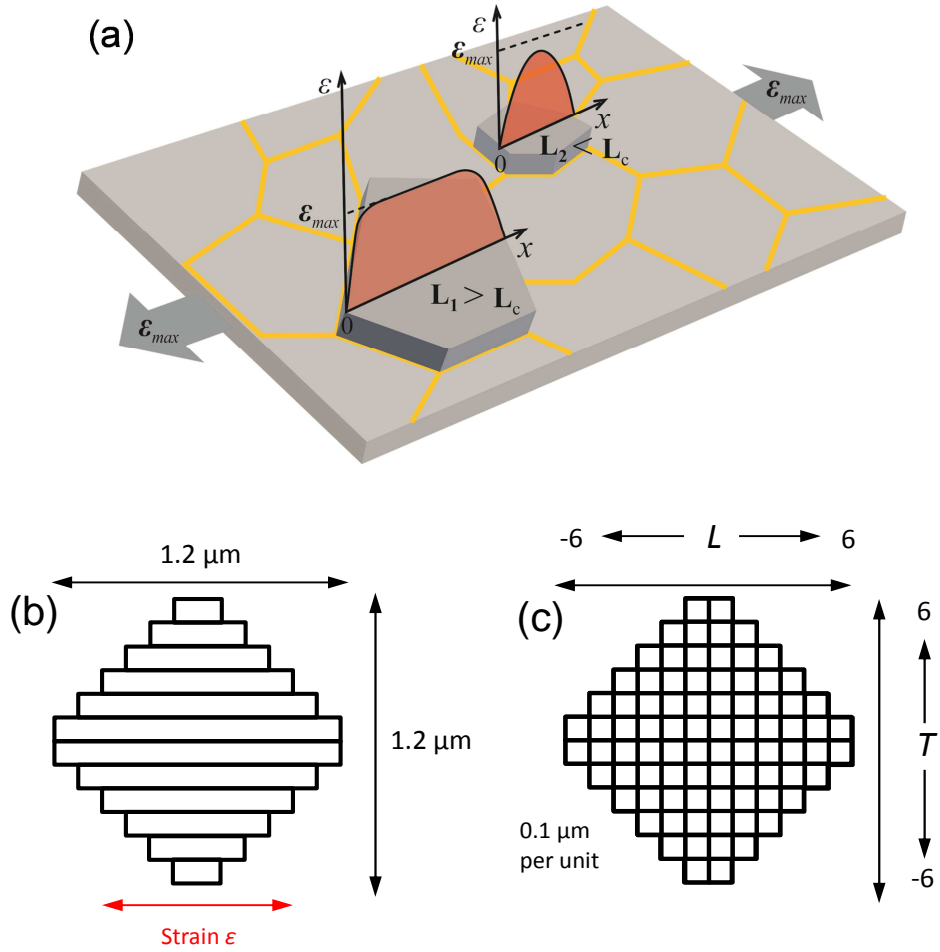

**Figure 4.** Schematic diagram (a) explaining the proposed stress transfer mechanism ($L_i$ is the length of the $i$-crystallite and $L_c$ the critical transfer length) and showing (b) the strips in the graphene islands and (c) the corresponding elementary units.

We now consider how stress transfer takes place from the PET substrate to graphene nanoribbon. It was shown using shear-lag theory[59, 61] that for an exfoliated monolayer graphene flake deformed on a polymer substrate, the strain in the graphene should be zero near the edges and increase towards the center of the flake such that:[33]

$$\varepsilon_r = \varepsilon_m \left[ 1 - \frac{\cosh\left(ns\frac{x}{l}\right)}{\cosh\left(\frac{ns}{2}\right)} \right] \quad (2)$$

where $n = \sqrt{\frac{2G_m}{E_g}\left(\frac{t_g}{t_m}\right)}$ (3)

and where $\varepsilon_m$ is the matrix strain and $\varepsilon_r$ is the real strain of graphene as a function of longitudinal position $x$ along the stress direction. In this case, $l$ is the length of the graphene nanoribbon along the stress direction, and $s$ (= $l/t_g$) is defined as the nanoribbon aspect ratio. $G_m$ and $E_g$ are the shear modulus of the matrix and the Young's modulus of graphene, respectively and $t_g$ and $t_m$ are the thickness of graphene and the elementary matrix, respectively. The parameter $ns$ is generally accepted to be a measure of stress transfer efficiency, being higher for better stress transfer efficiency, and also increasing proportionally with the size of the monolayer graphene flake or nanoribbon.[33] This theory implies a non-uniform strain in the graphene nanoribbons (and therefore the graphene islands) along the direction of axial stress, particularly when the nanoribbon is smaller than the 'critical length',[33] (twice of the distance it needs for strain to increase to the plateau value). This model of graphene islands isolated mechanically by wrinkles is analogous to the case of short fiber reinforcement in composites where there is no stress transfer across the fiber ends.[62]

The value of $ns$ is proportional to the length of the graphene nanoribbon $l$ (since $s \propto l$) thus both $ns$ and $l$ should be the function of the transverse position parameter $T$ (Figure 4(c)), i.e. $(ns)_T$ and $l_T$. Typically, the value of $ns$ is taken to be of the order of 2 for a graphene nanoribbon 1.2 μm ($T = \pm 1$) along the stress direction.[33] It may also vary with the efficiency of stress transfer between the substrate and graphene as will be discussed later. Due to the symmetry of the strain and laser spot intensity distributions, only the units with positions ($1 \leq L \leq 6$, $1 \leq T \leq 6$) have been considered here, and the distance between each unit (Figure 4(c)) is calculated through the unit center (i.e. the distance of unit (5,0) to the island center is calculated as (0.1×5)-0.05=0.45 μm). Based on this, eq 2 can be modified to determine the strain in each individual unit $\varepsilon_r(L,T)$ in Figure 4(c) as:

$$\varepsilon_r(L,T) = \varepsilon_m \left[ 1 - \frac{\cosh\left(\frac{(ns)_T}{l_T}(0.1L-0.05)\right)}{\cosh\left(\frac{(ns)_T}{2}\right)} \right] \qquad (4)$$

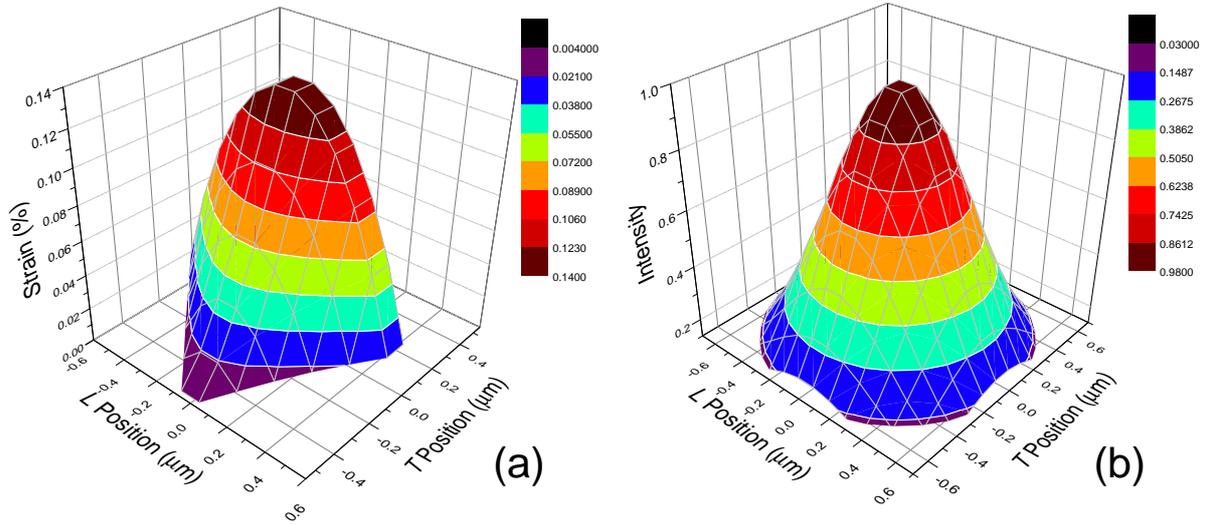

**Figure 5.** (a) Strain distribution in one island and (b) intensity distribution in the laser spot.

Figure 5(a) shows the predicted strain distribution within a ~1.2 μm diameter graphene island for a PET substrate strain of 0.4%. It can be seen that the strain is zero at the edges of the island and increases to a maximum of only 0.14% in the center of the island. This demonstrates clearly that the presence of the wrinkles reduces the efficiency of stress transfer to the graphene monolayer. In order to simulate the effect of deformation of the CVD graphene/PET upon the shift of the graphene 2D Raman band, however, both the non-uniform strain in the islands and any local variation in laser spot intensity have been taken into account. Figure 5(b) shows the intensity distribution within the laser spot calculated using Figure S1 and eq S3.

The 2D Raman band intensity collected from a unit ($L,T$) (Figure 4(c)) may be represented in the form of a Lorentzian function $I(\omega,L,T)$:[63]

$$I(\omega,L,T) \propto \frac{FWHM/2}{(\omega-\omega_{L,T})^2 + (FWHM/2)^2} \quad (5)$$

where $\omega_{L,T}$ is the position of the simulated band and $FWHM$ is its full width at half maximum. As constant band intensity persists even though the band position shifts so the maximum band intensity at $\omega_{L,T}$ is normalized to be unity by a factor of $FWHM/2$, without affecting its other band parameters. Combining eq 4, eq 5 and eq S3, the normalized intensity distribution for the 2D Raman band under strain for each unit ($L,T$) in the graphene island may be given as:

$$I(\omega,L,T) = \underbrace{\frac{\left[\left(FWHM_{2D} + \varepsilon_r(L,T)\left(\frac{dFWHM_{2D}}{d\varepsilon}\right)_{ref}\right)/2\right]^2}{\left(\omega-\omega_{L,T} - \varepsilon_r(L,T)\left(\frac{d\omega_{2D}}{d\varepsilon}\right)_{ref}\right)^2 + \left[\left(FWHM_{2D} + \varepsilon_r(L,T)\left(\frac{dFWHM_{2D}}{d\varepsilon}\right)_{ref}\right)/2\right]^2}}_{\text{Strain}} \cdot$$

$$\underbrace{\exp\left[-2\frac{(0.1L-0.05)^2 + (0.1T-0.05)^2}{r_0^2}\right]}_{\text{Intensity}} \quad (6)$$

This equation takes into account both the local strain in the unit and the local intensity of the laser spot. The 2D Raman band collected in the whole island $I_{\text{Total}}(\omega)$ can then be determined as the summation of the contribution of all the elementary units ($L,T$) in the island:

$$I_{\text{Total}}(\omega) = \sum_{T=1}^{6}\sum_{L=1}^{6} I(\omega,L,T) \quad (7)$$

For strain-free graphene, all the $\omega_{L,T}$ are taken as zero for simplicity, and $FWHM_{2D}$ is taken as the average value for a graphene monolayer flake of 27 cm$^{-1}$.[28, 64] The ideal values from exfoliated flat graphene flakes of $(d\omega_{2D}/d\varepsilon)_{ref} = -60$ cm$^{-1}$/% (eq 1) and $(dFWHM_{2D}/d\varepsilon)_{ref} = 12$ cm$^{-1}$/% can be used for 514 nm laser excitation, assuming a perfect interfacial adhesion within the graphene island.[34] As mentioned earlier, a typical value of $ns=2$ is used for a monolayer graphene of length 1.2 μm along the stress direction. The Raman 2D band for whole graphene island calculated using eq 7 is shown in Figure 6.

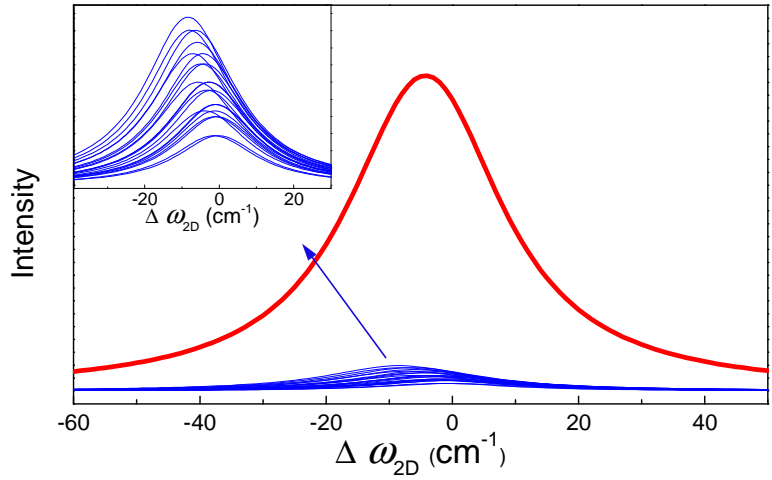

**Figure 6.** Simulated shift of Raman 2D band for each unit ($1 \leq L \leq 6$, $1 \leq T \leq 6$, blue curves) and the integrated 2D band for the whole graphene island (red curve).

Consequently, the unusual band shift and broadening behavior of the wrinkled graphene can be determined from the summation of the Raman scattering from the different elementary units under strain. The lower rate of band shift per unit strain is the result of the small size of the graphene islands limiting the maximum strain and the non-uniform strain distribution causing more band broadening than would otherwise be expected. These effects are not found in larger flat mechanically-exfoliated graphene flakes since the strain in them is reasonably uniform, except at the edges.[33, 47, 48]

In reality, the value of *ns* will vary depending upon the quality of interfacial stress transfer. With poor interfacial stress transfer, the *ns* value will be lower, leading to a less strained graphene island and a lower maximum strain at the island center. This less-strained graphene also results in different values of measured $d\omega_{2D}/d\varepsilon$ and $dFWHM_{2D}/d\varepsilon$. In this case, the variation of $\omega_{2D}$ and $FWHM_{2D}$ are predicted based on eq 6 and eq 7 using sets of $ns$ values corresponding to different levels of interfacial adhesion (Figure 7). It can be seen that both sets of experimental data fall close to the line for $ns$= 2 and 3, demonstrating that the stress transfer between the PET and the CVD graphene within the graphene island is fairly good, and comparable to the interface between exfoliated graphene and SU8/poly(methyl methacrylate).[33]

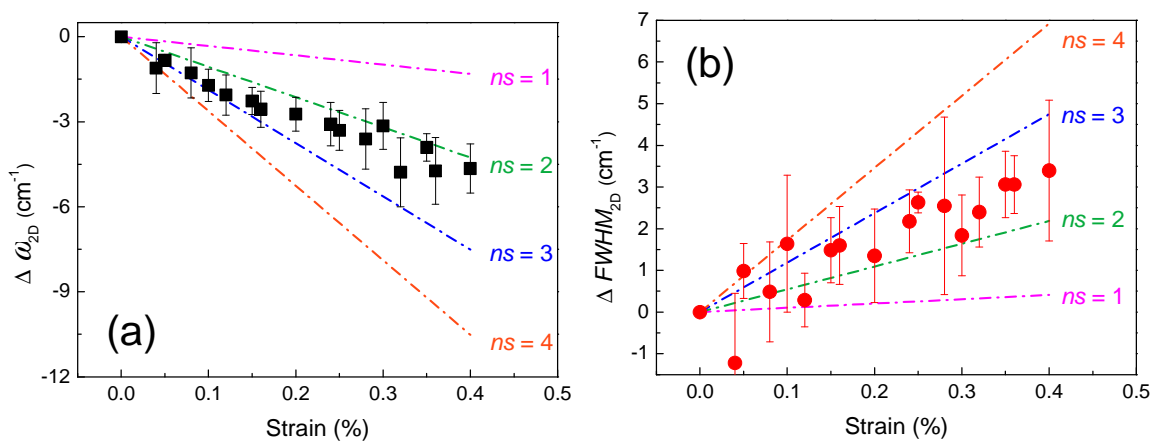

**Figure 7.** Predicted variation of (a) $\omega_{2D}$ (b) $FWHM_{2D}$ as the function of strain for different $ns$ values (used in eq 6 (colored lines). The data points are the experimental results from Figure 3.

The data in Figure.7 have been modeled assuming that the graphene islands and laser spots are approximately concentric. In reality, the Raman laser spot could be centered at any position relative to the graphene islands, either at their center, on their edges (Figure S2) or at any intermediate position. This behavior has been modeled in the Supporting Information. The variation of $\omega_{2D}$ and $FWHM_{2D}$ with strain for either the laser beam centered in the middle of the graphene island or at the edges have been calculated (Figure S3), again using $ns$ values of 2 and 3. It can be seen in the Supporting Information that the predicted variations of $\omega_{2D}$ and $FWHM_{2D}$ with strain vary significantly with the position of the laser spot. Nevertheless, it is possible to fit the simulated lines to the experimental data by choosing appropriate *ns* values for any laser spot position. The fact that both the band shift and band broadening data can be fitted using the same value of *ns* gives confidence in the validity of the model.

Finally, it is worth considering the effect of wrinkles upon the ability of graphene to reinforce nanocomposites. To a first approximation, the effective Young's modulus of the graphene scales with the Raman band shift rate per unit strain.[22, 65] Hence the band shift rate of the wrinkled graphene, being less than 25% of that of flat mechanically-exfoliated material, implies that it will have an effective Young's modulus of only around 250 GPa, as opposed to ~1TPa for flat graphene.[66] In this present study, however, the graphene has only one interface with the polymer substrate whereas there would be two for wrinkled graphene fully embedded in a polymer matrix.. We showed in an earlier paper[34] that the level of interfacial stress transfer between a polymer substrate and monolayer graphene is similar for both uncoated (one interface) and coated (two interfaces) model composite specimens. Hence, the findings in this present study can be related directly to the effect of wrinkles upon the deformation of graphene in bulk nanocomposites.

**CONCLUSIONS**

The deformation of wrinkled CVD graphene on PET substrate has been monitored through the use of Raman spectroscopy. It has been demonstrated that the unusual Raman band shift behavior observed is a result of the graphene microstructure, with mechanically-isolated graphene island of a comparable size to the Raman laser spot. By deconvoluting the Raman spectra obtained from the graphene networks, a model has been proposed to take account both the non-uniformity of local strain in the graphene microstructure and the intensity distribution in the laser spot. The good fit between the experimental data and the prediction confirms the appropriateness of this model, validating the use of this technique in estimating the effect of defects such as wrinkles on the performance of graphene-based devices. It also implies that when the characteristic dimensions of the microstructural units are of similar size to the spatial resolution of the Raman spectrometer laser spot, the conventional analysis has to be corrected to take into account both the structural non-uniformity and the resolution of the laser beam.

**MATERIALS AND METHODS**

The graphene for laser spot size determination in the Supporting Information was made by mechanical exfoliation and was then transferred to a PMMA substrate.[33] The CVD graphene was grown on copper using a conventional methane feedstock and was then transferred onto PET film as described in the Supporting Information. For the bending test, the CVD graphene/PET film was attached to PMMA beam by PMMA solution adhesive.

SEM images were obtained using a Philips XL30 FEGSEM. The sample surface was coated with gold before analysis. AFM images were obtained from the surfaces of the CVD graphene using a Dimension 3100 AFM (Bruker) in the tapping mode in conjunction with the 'TESPA' probe (Bruker).

Raman spectra were obtained using Renishaw 1000 spectrometers equipped with an Argon laser ($\lambda$ = 514 nm). The sample on the PMMA was deformed in a four-point bending rig, with the strain monitored using a resistance strain gauge attached to the PMMA beam adjacent to the CVD graphene/PET film. In all cases, the incident laser polarization is kept parallel to the strain. The simulation of Raman spectra was carried out using Wolfram Mathematica 9.

## ASSOCIATED CONTENT

**Supporting Information**. Transfer of the CVD graphene to the PET substrate. Determination of the lateral dimensions of the graphene islands. The effect of wrinkles upon stress transfer. Determination of the mean rates of 2D Raman band shift and band broadening. The estimation of the laser spot size and the local laser intensity calculation. The effect of laser spot position and the *ns* values upon $\omega_{2D}$ and $FWHM_{2D}$. This material is available free of charge via the Internet at http://pubs.acs.org.

## AUTHOR INFORMATION

**Corresponding Author**

*Robert J Young: robert.young@manchester.ac.uk

**Author Contributions**

The manuscript was written through contributions of all authors. All authors have given approval to the final version of the manuscript.

**Funding Sources**

The research leading to these results has received funding from the European Union Seventh Framework Programme under grant agreement n°604391 Graphene Flagship, EPSRC (award no. EP/I023879/1) and AFOSR/EOARD (award no. FA8655-12-1-2058). The Patras group also acknowledges the support of the ERC Advanced Grant "Tailor Graphene" (no: 321124).

## ACKNOWLEDGMENT

One of the authors (Z-L. L.) is grateful to the China Scholarship Council for financial support.

TOC

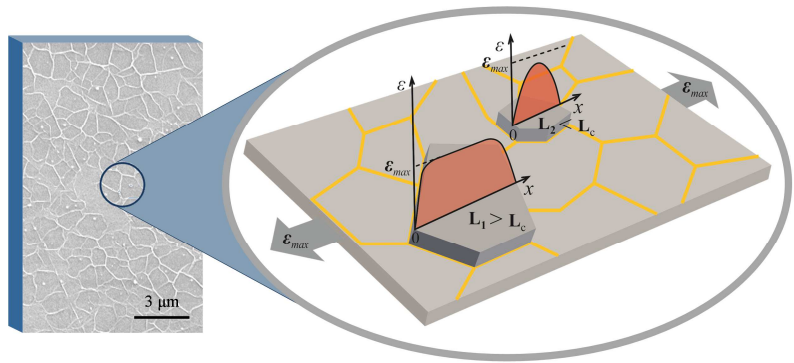

# Deformation of Wrinkled Graphene

# (Supporting Information)

Zheling Li[1], Ian A. Kinloch[1], Robert J. Young[1]*, Kostya S. Novoselov[2], George Anagnostopoulos[3], John Parthenios[3], Costas Galiotis[3,4], Konstantinos Papagelis[3,5], Ching-Yu Lu[6], Liam Britnell[6]

[1]School of Materials and [2]School of Physics and Astronomy, University of Manchester, Oxford Road, Manchester, M13 9PL, UK

[3]Institute of Chemical Engineering Sciences, Foundation for Research and Technology – Hellas (FORTH/ ICE-HT), P.O. Box 1414, Patras 265 04, Greece,

[4]Department of Chemical Engineering and [5]Department of Materials Science, University of Patras, Patras 26504, Greece

[6]BGT Materials Ltd, Photon Science Institute, University of Manchester, Oxford Road, Manchester, M13 9PL, UK

1. **Transfer of the CVD graphene to the PET substrate**

The CVD graphene was transferred to PET using the well-known technique of chemical etching of copper in ferric chloride. The graphene/copper had a poly(methyl methacrylate) (PMMA) coating applied to the surface to act as a support during transfer. This was allowed to dry at room temperature. The backside graphene was removed by oxygen plasma. The PMMA/graphene/copper foil was then placed in ferric chloride until all copper was dissolved. The film was then subsequently transferred to three baths of DI water. The PMMA/graphene film was then fished from the last DI bath with a clean PET film and allowed to dry overnight. The PMMA/graphene/PET film was then soaked in acetone to remove the PMMA. The graphene/PET film was rinsed in IPA and blown dry with nitrogen as a final cleaning step.

2. **Determination of the lateral dimensions of the graphene islands**

The lateral dimension of each graphene island was estimated by averaging the length of two crossed lines across the island, as shown in Figure S1. Over 500 islands were measured in this way to give the statistical distribution in Figure 1(c). A typical example of this measurement is shown in Figure S1(b).

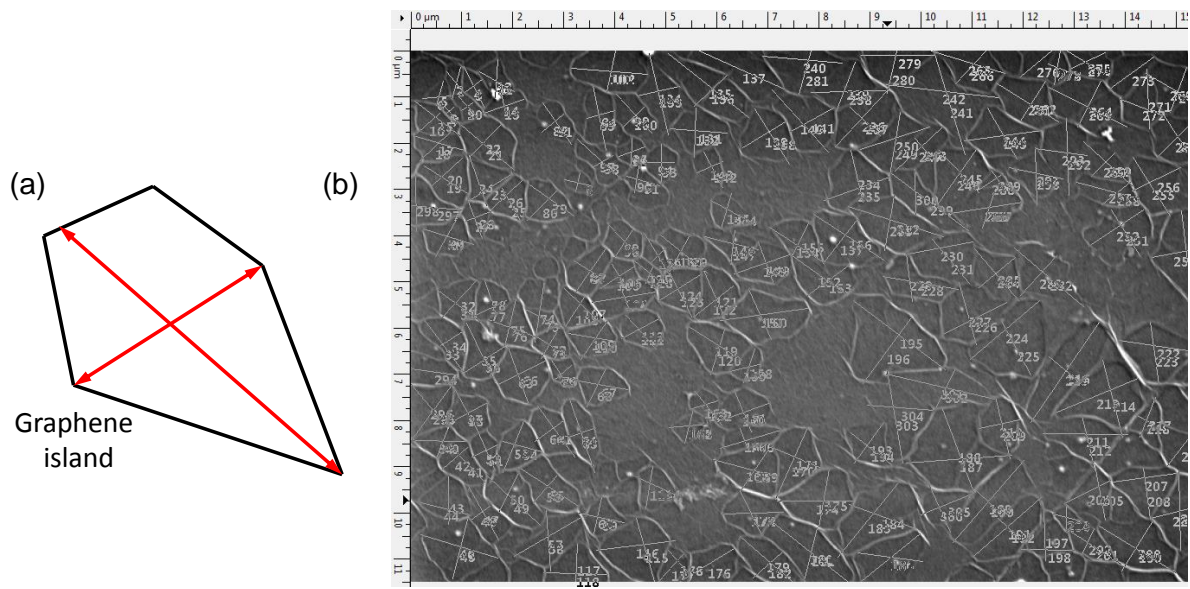

**Figure S1.** (a) Method employed to measure the lateral dimension of the graphene islands from SEM micrographs. The black line is the edge of graphene island, and its lateral dimension is measured by averaging the length of the two red lines across it. (b) A typical SEM micrograph showing the individual measurements.

**3, The effect of wrinkles upon stress transfer**

The AFM height scan in Figure 1(d) in the manuscript clearly shows that the wrinkles stick up above the PET substrate. As shown before for a different system, when the wrinkles form by a similar mechanism, the upper layer stick up and leave a hollowed region between the top and bottom layer.[1] Similarly in our situation, there will be a hollow region within the wrinkles in which there can be no stress transfer giving rise to the mechanically-isolated graphene islands and mechanically-free edges within the wrinkles because of the absence of the interface with the substrate, as shown schematically in Figure S2. Deformation will also lead to straightening of the wrinkles. It should be noted that the diagrams in Figure S2 are not to scale – the size of the islands (1.2 μm) is much larger than the height of the wrinkles (~20 nm).

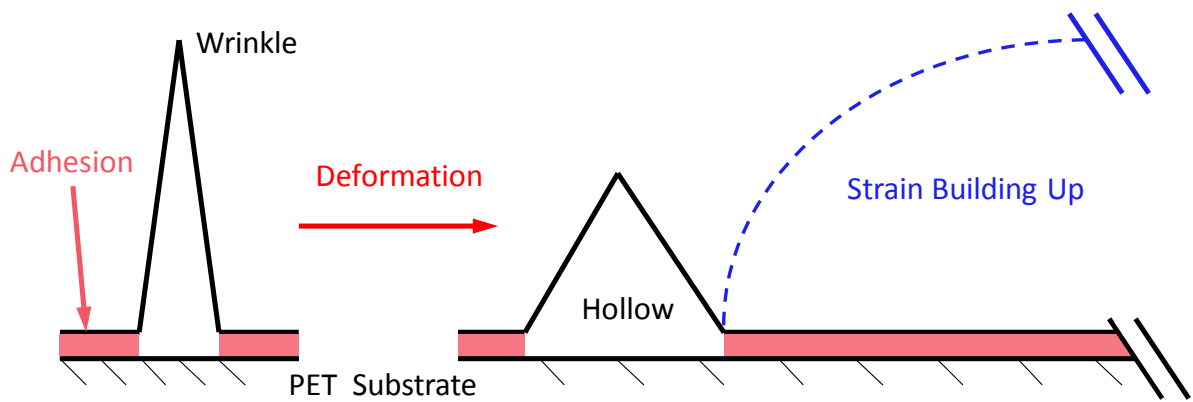

**Figure S2.** Schematic diagram of the deformation of the wrinkled structure (not to scale).

## 4. Determination of the mean rates of 2D Raman band shift and band broadening

The mean size of the graphene islands was fixed by the process used to produce the CVD graphene/PET material although Figure 1(c) shows that there was a wide distribution of sizes. Although we were not able to modulate the mean value, we were able to investigate the effect of variation in the lateral dimensions indirectly by undertaking measurements in random position on specimens as shown in Figure S3. It can be seen that there are wide variations in the rates of both band shift and band broadening for the 2D band. The 8 sets of data are more scattered than for similar measurement on flat mechanically-exfoliated graphene[2] and the difference in behavior is most likely due to the variations of lateral dimensions in the regions chosen at random.

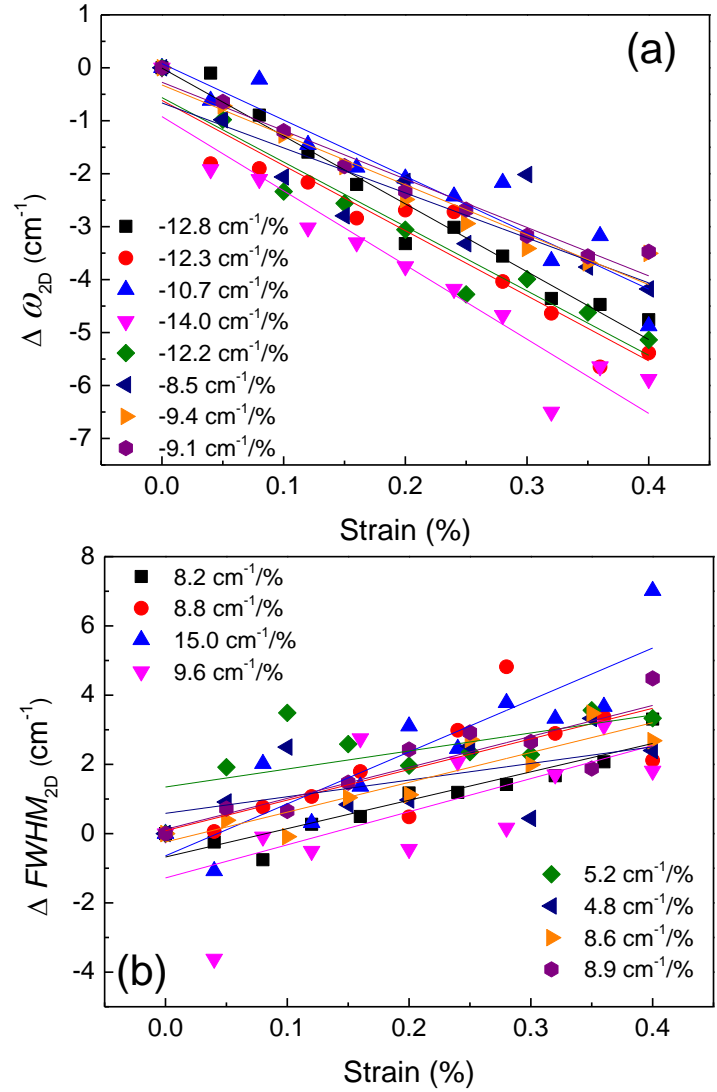

**Figure S3.** The variation of (a) $\omega_{2D}$ (b) $FWHM_{2D}$ as the function of strain in each test, with the $d\omega_{2D}/d\varepsilon$ and $dFWHM_{2D}/d\varepsilon$ also indicated.

The Raman spectra of a typical CVD graphene/PET specimen at different strain level are in shown in Figure S4.

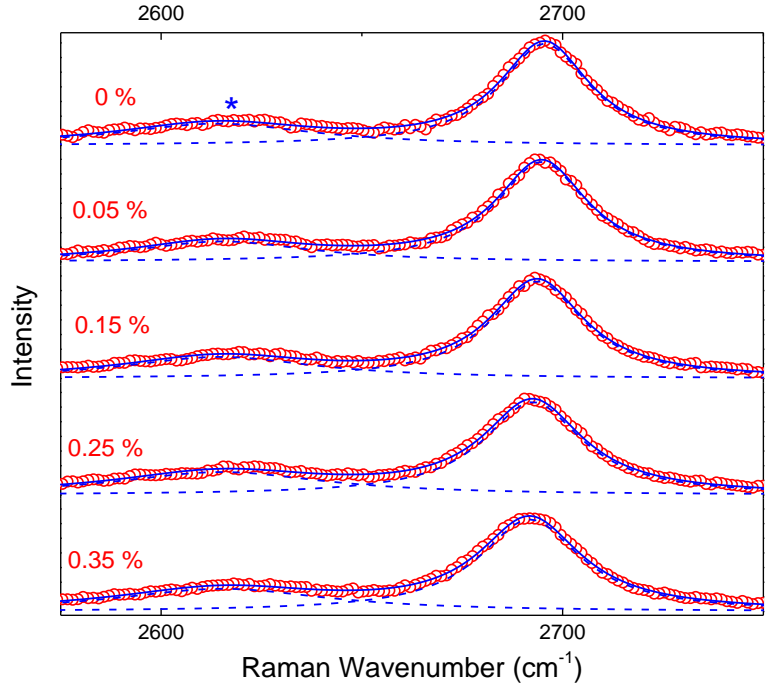

**Figure S4.** Raman spectra of the wrinkled CVD graphene on PET at different strain level. The red dots and blue curves denote the experimental spectra and curve fittings, respectively. The blue '*' indicates the Raman band from PET substrate.

## 5. The estimation of the laser spot size and the local laser intensity calculation

It can be assumed that the laser intensity $I(r)$ within the spot of a Gaussian laser beam follows the Gaussian distribution:[3]

$$I(r) = \exp\left(-2\frac{r^2}{r_0^2}\right) \qquad (S1)$$

where $r$ is the distance to the laser spot center, and $r_0$ is the radius of the laser beam, defined as the radius of the plane where $I(r)$ decreases to $1/e^2$ of its maximum value.

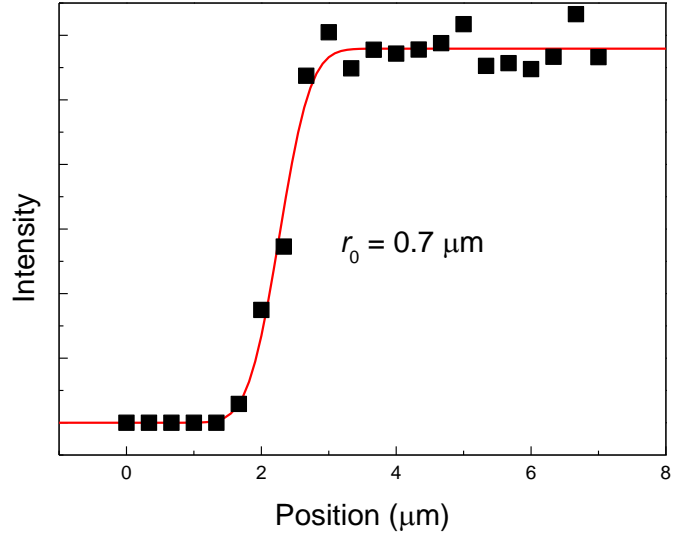

**Figure S5.** Variation of *I* when the laser beam is moving across a well-defined graphene edge.

The spot size of the Gaussian beam laser[3] was estimated by moving the laser spot inwards from outside a mechanical exfoliated monolayer graphene towards its center, and the 2D band intensity *I* can be fitted as the function of laser position $x_l$ (Figure S5):[4]

$$I = \sqrt{\frac{\pi}{8}} A r_0 \left(1 + \mathrm{erf}\left(\frac{\sqrt{2}(x_l - x_0)}{r_0}\right)\right) \quad (S2)$$

where $x_0$ is the graphene edge location and *A* is the amplitude. For regions where the 2D was not resolvable, *I* was set as zero. By fitting *I* with eq S2, the radius of the laser beam $r_0$ is obtained as ~0.7 μm. Thus diameter of the laser spot size is estimated to be 1.4 μm.[5]

If the distance between each unit (Figure 4(c)) is then calculated through the unit center, the local laser intensity at unit (*L*,*T*), $I_{\mathrm{laser}}(L,T)$ is given by modification of eq S1 as:

$$I_{\mathrm{laser}}(L,T) = \exp\left[-2 \cdot \frac{(0.1L - 0.05)^2 + (0.1T - 0.05)^2}{r_0^2}\right] \quad (S3)$$

## 6. The effect of laser spot position and the *ns* values upon $\omega_{2D}$ and $FWHM_{2D}$

The effect of the exact laser spot position and the *ns* values to the variation of $\omega_{2D}$ and $FWHM_{2D}$ are also considered. The laser spot is approximated to a square and Figure S6 shows the situation where the laser spot is centred at the wrinkles. The overlapped region of graphene islands and laser spot contributes to the calculated Raman spectra, and only the region marked by dashed red lines was taken into calculation due to its symmetrical geometry.

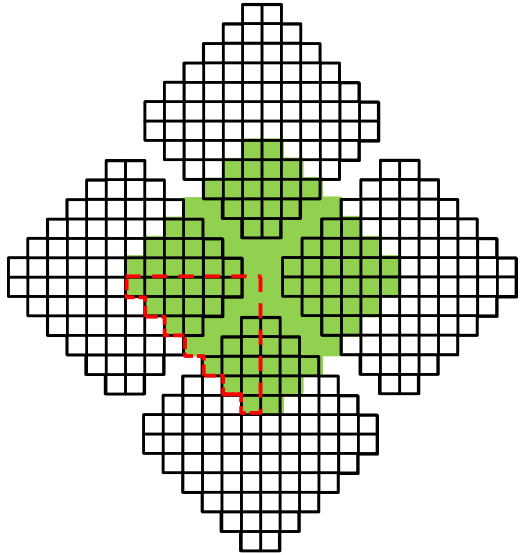

**Figure S6.** Schematic diagram of the laser spot (approximated to a square) centered at the wrinkles (the gaps) between the graphene islands (black squares). The overlapped region of graphene islands and laser spot (green square) contributes to the calculated Raman spectra.

The effect of the exact laser spot position and the *ns* values to the variation of $\omega_{2D}$ and $FWHM_{2D}$ are shown in Figure S7.

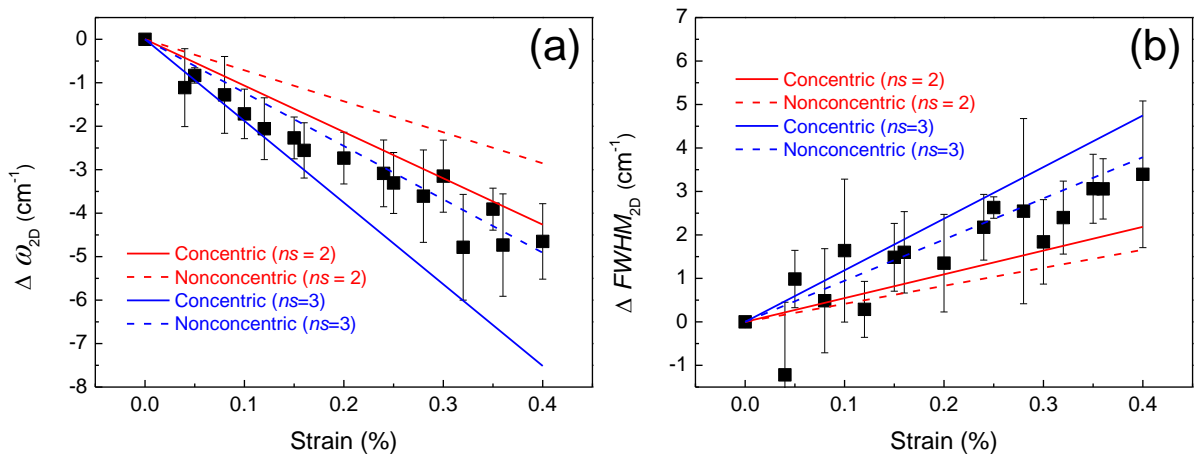

**Figure S7.** Predicted variation of (a) $\omega_{2D}$ (b) $FWHM_{2D}$ as the function of strain when graphene island and laser spot are concentric (solid lines) and nonconcentric (dashed lines), with an *ns* value of 2 (red lines) and 3 (blue lines). Black squares are the experimental results.

It can be seen that the experimental data can be fitted by choosing the appropriate combination of spot position (concentric or nonconcentric) and *ns* values.